\documentclass[journal]{IEEEtran}
\usepackage{cite}

\ifCLASSINFOpdf
\else
\fi

\usepackage{graphicx}
\usepackage{stfloats}

\usepackage{booktabs}
\usepackage{array}
\usepackage{epstopdf}

\usepackage[mathlines]{lineno}

\usepackage{authblk}
\usepackage{subfigure}
\usepackage{fancyhdr}		

\usepackage{amsmath, xparse}
\usepackage{multirow}

\usepackage{color}

\usepackage{relsize}

\usepackage{hyperref}
\hypersetup{ colorlinks,
    citecolor=blue,
    filecolor=blue,
    linkcolor=blue,
    urlcolor=blue
}
\usepackage{xcolor}
\definecolor{gray}{rgb}{0.6,0.6,0.6}
\definecolor{red}{rgb}{0.85,0,0}
\definecolor{green}{rgb}{0,0.85,0}
\definecolor{blue}{rgb}{0,0,0.85}
\definecolor{beige}{rgb}{0.92,0.87,0.78}
\usepackage[all]{hypcap}

\begin{document}

\title{An Explainable Non-local Network for COVID-19 Diagnosis}

\author{Jingfu~Yang,
        Peng~Huang,
        Jing Hu, 
        Shu~Hu,
        Siwei Lyu,
        Xin~Wang,
        Jun~Guo$^{*}$,
        Xi~Wu$^{*}$
\thanks{Jingfu Yang, Jing Hu, and Xi Wu are with the Chengdu University of Information Technology, China. e-mail:yangjf7@gmail.com, \{lesley123, wuxi\}@cuit.edu.cn}
\thanks{Peng~Huang is with the school of Computing and Artificial Intelligence, Southwest Jiaotong University. e-mail:huangpeng@my.swjtu.edu.cn}
\thanks{Xin Wang is with the University at Albany, State University of New York. e-mail:xwang56@albany.edu}
\thanks{Shu Hu is with the department of Computer and Information Technology, Purdue University. e-mail:hu968@purdue.edu}
\thanks{Siwei Lyu is with the University at Buffalo, SUNY, USA. e-mail: siweilyu@buffalo.edu}
\thanks{Jun Guo is with the department of Critical Care Medicine, West China Hospital, Sichuan University. e-mail:guojun@wchscu.cn}
\thanks{$^{*}$ Corresponding authors.}
}

%



\maketitle

\begin{abstract}
The convolutional neural network has achieved excellent results in the automatic classification of medical images. In this study, we propose a novel deep residual 3D attention non-local network (NL-RAN) to classify CT images included COVID-19, common pneumonia, and normal to perform rapid and explainable COVID-19  diagnosis. We built a deep residual 3D attention non-local network that could achieve end‐to‐end training. The network is embedded with a nonlocal module to capture global information, while a 3D attention module is embedded to focus on the details of the lesion so that it can directly analyze the 3D lung CT and output the classification results. The output of the attention module can be used as a heat map to increase the interpretability of the model. 4079 3D CT scans were included in this study. Each scan had a unique label (novel coronavirus pneumonia, common pneumonia, and normal ). The CT scans cohort was randomly split into a training set of 3263 scans, a validation set of 408 scans, and a testing set of 408 scans. And compare with existing mainstream classification methods, such as CovNet, CBAM, ResNet, etc. Simultaneously compare the visualization results with visualization methods such as class activate mapping (CAM). Model performance was evaluated using the Area Under the ROC Curve(AUC), precision, and F1-score. The NL-RAN achieved the AUC of 0.9903, the precision of 0.9473, and the F1-score of 0.9462, surpass all the classification methods compared. The heat map output by the attention module is also clearer than the heat map output by CAM. Our experimental results indicate that our proposed method performs significantly better than existing methods. In addition, the first attention module outputs a heat map containing detailed outline information to increase the interpretability of the model. Our experiments indicate that the inference of our model is fast. It can provide real-time assistance with diagnosis.
\end{abstract}


%
\IEEEpeerreviewmaketitle

\section{Introduction}
%
%
%
%
\IEEEPARstart{N}{ovel} Coronavirus 2019 (COVID-19), which was discovered due to viral pneumonia cases in Wuhan in 2019, was named by WHO on January 12, 2020. On May 5, 2023, the WHO announced that COVID-19 no longer constitutes a Public Health Emergency of International Concern. Prior to this announcement, the total number of reported COVID-19 deaths to the WHO was nearly 7 million, and health systems worldwide were severely disrupted. Research indicates that the earlier COVID-19 is detected and targeted treatment is administered, the less damage it inflicts on the patient's physical functions~\cite{Fang2024}.

The most common symptoms of COVID-19 patients include fever, cough, and shortness of breath. Severe COVID-19 patients typically suffer from pneumonia. Computed Tomography (CT) imaging plays a critical role in the detection of manifestations in the lung associated with COVID-19, where the classification of infection lesions from CT scans is important for the quantitative measurement of disease progression, accurate diagnosis, and follow-up assessment. A manual classification of infection lesions from 3D volumes is labor-intensive, time-consuming, and potentially suffers from inaccurate or inconsistent diagnosis due to variations of observers' training and experiences, automatic classification of infection lesions is highly desirable in clinical practice.

\begin{figure}[t]
\centering
\includegraphics[width=\columnwidth]{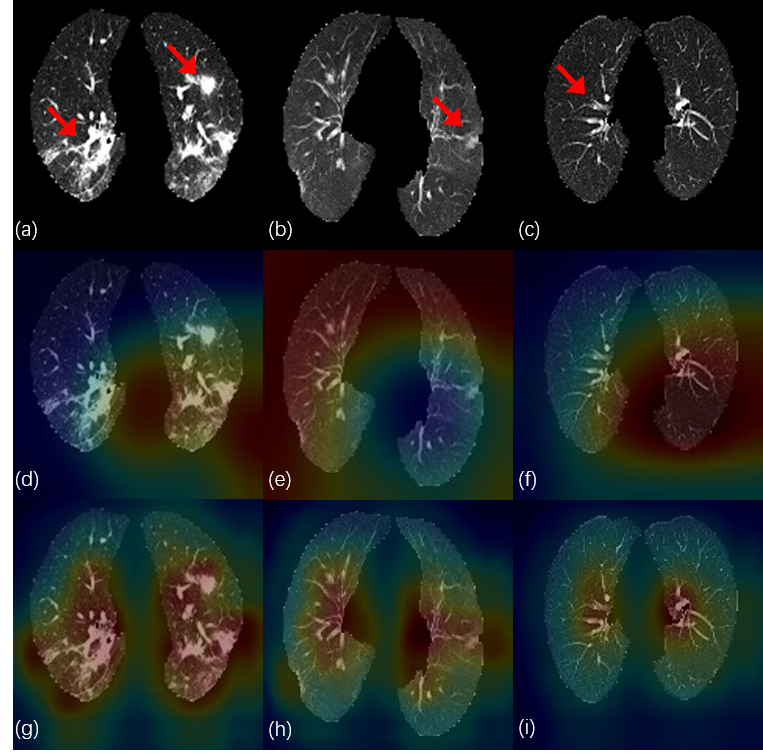}
\caption{Complex appearances of pneumonia lesions in CT scans of COVID-19 patients. (a-c) are COVID-19, common pneumonia, and normal, respectively, where red arrows highlight some lesions or suspected lesions.  
{There is a high similarity between the lesion area in (b) and the suspected lesion area in (c).} (d-f) are temperature maps inferred by the CAM. (c), (f) and (i) are normal lungs, but some tissues have suspected lesions that are captured by the neural network. (g-i) is the temperature map inferred by our model, which is more accurate in detail and positioning than CAM.}
\label{problem}
\end{figure}

Some existing studies utilize common computer vision techniques to analyze and process signs of COVID-19 in chest CT or X-ray images~\cite{JYOTI2023106331,GOEL2022105587, XU2023106417}. In clinical examinations, CT scans are crucial for the diagnosis and treatment decisions of COVID-19. However, automatically classifying pneumonia lesions from CT scans remains challenging due to several reasons. First, infection lesions have a variety of complex appearances such as Ground-Glass Opacity (GGO), reticulation, consolidation, etc. Second, the sizes and positions of pneumonia lesions vary significantly at different stages of infection and among different patients. Third, lesions have irregular shapes and ambiguous boundaries. Some lesion patterns such as GGO have low contrast with surrounding regions, and some lesions are indistinguishable from normal, as shown in subfigures (a-c) of Fig.~\ref{problem}. These challenges make it difficult to automatically classify pneumonia lesions. 

Recent efforts have been directed to the automatic classification of pneumonia lesions from CT scans. Some \cite{li2020artificial, hemdan2020covidx, huang2024robustly, tsai2024uu, lin2024robust} apply off-the-shelf CNN models such as ResNet for classification and adopt a 2D method by ignoring the existence of {information between slices}. However, using a 2D model to process CT leads to loss of information in a certain dimension of the three-dimensional volume and results in a decrease in the classification accuracy. Others use binary classification models for training, which ignores categories of pneumonia lesions other than COVID-19 and thus leads to mistaking other pneumonia lesions as COVID-19. In order to increase the sizes of receptive fields, some researchers stack convolutional layers \cite{he2016deep}\cite{wang2017residual}. This may result in contour details of lesions being largely ignored, making lesions difficult to be recognized. 

Furthermore, current detection systems lack enough interpretability. The problem can be mitigated by using interpretable methods \cite{li2018tell,fong2017interpretable,petsiuk2018rise,ribeiro2016should} include class activation mapping (CAM) \cite{zhou2016learning,naidu2020ss,desai2020ablation} and Grad-CAM \cite{selvaraju2017grad,chattopadhay2018grad,omeiza2019smooth}. For example, COVIDX-Net \cite{hemdan2020covidx}, CovNet \cite{li2020artificial}, and other methods \cite{wang2020deep}\cite{gozes2020rapid} have realized rapid classification of pneumonia lesions. However, the realization of CAM/Grad-CAM depends on the modification of the model structure and the output is difficult to contain contour information. As shown in (d) in Fig. \ref{problem}, some important lesions are missed in the CAM based heat map, in (e,f) many irrelevant areas are highlighted, such as the background. 




Meanwhile, learning-based attention mechanism \cite{vaswani2017attention,choi2016retain,chu2017online,yan2019stat,wu2024efficient,hu2023attention,lu2023attention,guo2022robust} has been widely investigated for focusing on specific details. Some novel attention modules \cite{zhang2018image,tang2019multi,tong2020channel,panboonyuen2019semantic} are attracting increasing attention as they do not require a specific network structure and can be easily combined with typical training procedures. For example, channel attention and spatial attention have been shown to be robust for image classification \cite{hu2018squeeze}\cite{woo2018cbam}. 
However, the size of receptive fields in recently works \cite{he2016deep}
are typically increased by stacking attention modules, which increases the complexity and parameters of the model and may degrade the performance of the model if the stacking is not designed carefully, especially for the large size 3D biomedical images \cite{almeida20213d,luu2021effciently,fu20213d,su2020dir,younes2021machine,tan2021automatic,lei2021breast,jun2020automated}. 


Moreover, for our COVID-19 study, the size and shape of the lesions have a large variance, and the locations of lesions are also random, may occur anywhere in the lung region. 

Therefore, it is necessary to integrate context information by capturing the remote dependency of the pixel. Integrating contextual information by capturing the remote dependence of pixel mode can establish a global relationship among the lesions and improve the accuracy of classification.



To this end, we propose a novel network to improve classification performance and interpretability jointly by learning from long-range dependencies. 
In our method, a 3D residual network is selected as the backbone \cite{liu2017two,maturana2015voxnet,garcia2016pointnet,huang2016point}.
At the same time, a 3D attention module was embedded in the backbone to keep the model focused on the details of the lesion. Finally, in order for the model to learn the relationship between lesions and the whole image, a non-local module is embedded in the model.
Specifically, for each pixel, our non-local module can obtain context information by calculating the relationship between any pixel point and other pixel points. By further cycling, each pixel eventually captures the long-range dependencies of all pixels. 
Experimental results on a large-scale public dataset demonstrate that our proposed network achieves the state-of-the-art  COVID-19 classification performance. In addition, to enable interpretability, we use the attention map from the attention module of the model as a thermogram for visual analysis. It provides a more accurate visualization of infected areas and improves the model interpretability of existing methods. As shown in (g-i) in Fig. \ref{problem}, our attention map localized all the lesions accurately with more tight boundaries.


The main contributions of this paper are as follows: (1) We propose a 3D framework based on a 3D attentional convolutional network to predict COVID-19  from  CT images. It stacks attention modules to focus on the detail of infected areas. (2) We embed a non-local module to enable the model to focus on the global information of the input, including information among slices and among pixels. The combination of the attention mechanism and non-local enables our model to focus on both image details and global information of the input, improving the COVID-19 classification performance. (3) We output the attention map as the heat map to provide visualization analysis of our model, which provides a more accurate visualization of infected areas in an image and better model interpretability than existing methods.  


 

\section{Related Work}

\noindent \textbf{COVID-19 Classification. } Yazdani et al.  developed a deep learning framework, named CT-Net, to predict COVID-19 from CT images \cite{yazdani2020covid}. It uses an attentional convolution network to focus on infected areas of the chest, enabling it to perform a more accurate prediction. Tang et al.~\cite{qian2020intelligent} used neural networks and digital image processing technologies to design a lightweight COVID-19 classification model based on the attention mechanism within and between slices. On this basis, they developed a covid-19 intelligent diagnosis system of early screening, lesion evaluation, and COVID-19 diagnosis. 
The system integrates segmentation and pixel distribution histograms of lungs and lesions. A framework is proposed in \cite{xu2020deep} that uses a 3D deep learning model to separate candidate infection regions from the pulmonary CT image set and uses a location-attention classification model to categorize the separated images into one of the COVID-19, influenza a viral pneumonia (IAVP), and irrelevant to infection (ITI) groups, together with corresponding confidence scores. A 3D deep learning framework, called CovNet, is proposed in \cite{li2020artificial} for detecting COVID-19 by extracting 2D local and 3D global representative features. CovNet uses RestNet50 as the backbone. It takes a series of CT slices as input to generate their features, which are then combined by a maximum pooling merge operation. The final feature map is fed to the fully connected layer, and the softmax activation function generates a probability score for each type (COVID-19, common pneumonia, and normal). Zhu et al. proposed the OPT-CO method~\cite{ZHU2024121141}, which utilizes a stochastic configuration network to optimize a pre-trained Transformer model for efficient COVID-19 classification. This method enhances model performance under extended parameter settings.



\noindent \textbf{Class Activation Mapping. }
A convolutional neural network not only has strong image processing and classification ability but also can locate key parts in an image. The class activation mapping (CAM) method can use this positioning information to generate a heat map, which indicates areas that have a greater impact on the output. Researchers often use this method to improve the interpretability of their models \cite{wu2021jcs}. However, using this method requires replacing the full connection layer of the model with the global average pooling layer (GAP). GAP directly sums up the spatial information of the feature layer and integrates the whole spatial information. The resulting network is more robust to the spatial variation of the input. The authors in \cite{lin2013network} propose to use GAP instead of the full connection layer. One of the major weaknesses of the full connection layer is the large number of parameters, especially when the full connection layer is connected with the last convolutional layer. Using GAP instead of a full connection layer is effective in reducing parameters and the risk of overfitting.

\noindent \textbf{Attention Mechanism. } In terms of image classification, Residual Attention Network (RAN)~\cite{wang2017residual} successfully combines an extremely deep convolutional neural network with the human visual attention mechanism and achieves far better accuracy and parameter efficiency than previous network structures. Compared with ResNet-152, RAN can achieve better performance with only 52$\%$ parameters and 56$\%$ flops, which indicates that the attention mechanism can significantly improve the network performance and reduce the complexity of the model. Although the attention module plays an important role in classification, a simple superposition of attention modules, as pointed out in \cite{wang2017residual}, causes a decline in the performance of the model. To address this problem, the authors of \cite{wang2017residual} put forward the attention map and combine it with the original feature map. This combination allows attention modules stacked like a convolution module and improves the performance.

The innovation of SENet \cite{hu2018squeeze} is to focus on the relationship between channels in hopes that the model can automatically learn the importance of different channel information. SENet uses ResNet as the backbone, embeds the channel attention mechanism into the residual block, and proves the effectiveness of the channel attention mechanism through experiments on the ImageNet dataset. This channel attention mechanism enables the model to pay more attention to the channel with the largest amount of information while suppressing unimportant channel characteristics. It won the champion of the 2017 ImageNet classification competition. Similarly, in the super-resolution task, RCAN \cite{zhang2018image} introduces the attention mechanism to treat different channels differently in order to improve the representing ability of the network. In our method, we draw on the channel attention mechanism and embed it in our model.
The core of the work CBAM \cite{woo2018cbam} is the combination of channel attention and spatial attention \cite{chen2017sca}, which has achieved good results in image classification. It refines the attention-based features into two different modules, embeds ResNet's residual block, and combines both channel attention and spatial attention to achieve a significant performance improvement. In our method, we use its spatial attention for reference and embed it into our model.

\noindent \textbf{Non-local Learning Methods. }Non-local \cite{wang2018non} proposes a building block for non-local operations that works on the long-range dependencies for video and images. Both computer vision and natural language processing tasks need to capture long-range dependencies. In sequential tasks, the RNN operation is a major means to capture long-range dependencies. While in CNN, multiple convolution modules are stacked to form a large receptive field: both convolution and loop operators are local operations in space and time, and capturing long-range dependencies is obtained through repeated stacking and backpropagation. However, such an approach has two problems. First, the efficiency of capturing long-range dependencies is very low. Second, due to the depth of the network, a careful design of modules and gradients is required. Non-local addresses these problems. It is computationally efficient, requiring fewer stack layers and maintaining constant input and output scales to achieve the same effect. It can be easily embedded into current network architectures. Non-local can also be used as a component and combined with other network structures. 


\begin{figure*}[t]
\centering
\centerline{\includegraphics[width=\textwidth]{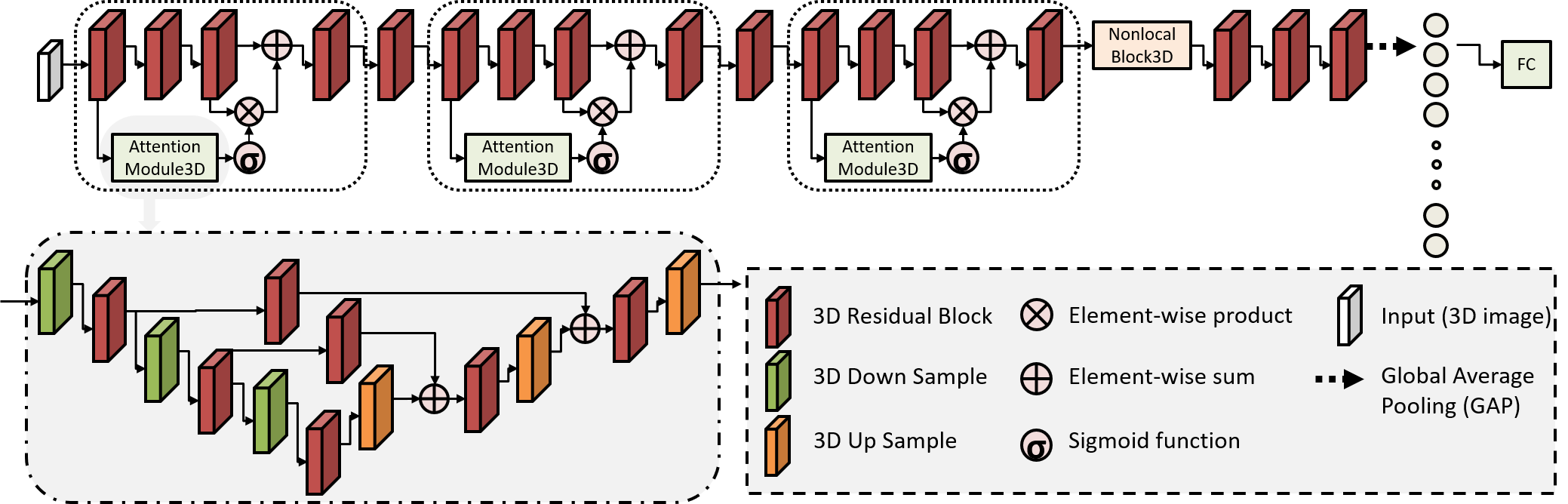}}
\caption{The architecture of our proposed NL-RAN, which consists of three mixed attention modules connected to the ResNet. After the last 3D attention module, a non-local module is embedded to enhance the receptive field, so that the model pays attention to both the details and the global information. }
\label{fig:NL-RAN}
\vspace{0.5cm}
\end{figure*}
\section{Non-local Residual Attention Network}
\subsection{Overview}
In this section, we present the non-local residual attention network (NL-RAN) to solve the pneumonia lesions classification problem. Fig. \ref{fig:NL-RAN} shows the network architecture of the proposed method. NL-RAN has composed of a backbone network and two key modules, the residual attention module and the non-local module.  We first introduce our residual attention module and the Non-local module, we then describe the network structures and loss functions, and finally we introduce how we obtain the attention maps.

\subsection{Residual Attention Module}

Fig. \ref{fig:getAttMap} shows the architecture of the proposed 3D attention module. It is composed of a 3D bottom-up top-down feedforward structure with multiple downsamplings and upsamplings. First, the input feature map is downsampled once. The result is sent to two branches through a residual block, one branch is a jump connection, and the other is downsampling and a residual block. After two downsamplings, the upsampling operation is performed. The upsampling result is superimposed with the output of the jump connection branch, followed by another upsampling operation. After two upsamplings, a weight matrix, known as attention map \cite{wang2017residual}, is used. This process is similar to the location learning process in the segmentation task. In addition, the attention map can be output as a 
heat map to increase the interpretability of the proposed model.

The residual attention module is expressed as:
\begin{equation}
H_{i, c}(x)=\left(1+M_{i, c}(x)\right) * F_{i, c}(x)
\end{equation}
where $F_{i, c}(x)$ is the feature map of the main trunk, and $M_{i, c}(x)$ is the attention map of the same size as the output of the main trunk, with $i$ ranging over all spatial positions and $c$ standing for a channel.

The attention module is divided into mixed attention, channel attention, and spatial attention by the activation function. The mixed attention uses sigmoid for each channel and each spatial location:
\begin{equation}
f_{1}\left(x_{i, c}\right)=\frac{1}{1+\exp \left(-x_{i, c}\right)}
\end{equation}
The attention map obtained by the attention module adjusts the weight relationship in spatial and channel. Direct activation can preserve attention information in both spatial and channel.


{Spatial and channels are from the feature map of convolution operation. Generally speaking, using multiple convolution kernels to convolute the image will get the corresponding feature map. The spatial size of the feature map is the resolution size of the image after the convolution operation, and the channel size is the number of convolution kernels, so the size of the feature map is the resolution size multiplied by the number of channels. The spatial information refers to the internal relationship of the resolution size of a channel in the feature map, and channel information refers to the relationship between channels in the feature map. Because the feature map obtained by convolution operation has confused spatial information and channel information. If we want to use attention mechanism in spatial or channel, we can only use some operations to remove the relationship between channels to retain spatial information, or remove the spatial relationship within the channel to retain channel information. For example, L2 normalization is performed on all channels for each spatial location to delete spatial information. Or, normalization is performed on the feature map of each channel, followed by a Sigmoid transformation to remove the connections between channels. }

The channel attention performs the L2 normalization for each spatial position of the attention map in all channels, with the spatial information removed from the attention map while the relationship among channels preserved:
\begin{equation}
f_{2}\left(x_{i, c}\right)=\frac{x_{i, c}}{\left\|x_{i}\right\|}
\end{equation}
Among them, the $x_{i}$ represents all channels of each spatial position in the attention map. the $x_{i, c}$ represents each channel of each spatial position in the attention map.

The spatial attention uses the following function $f_{3}$ to perform normalization within the feature map from each channel, with the relationship among channels removed and the spatial information retained:
\begin{equation}
f_{3}(x_{i, c})=\frac{1}{1+\exp (-(x_{i, c}-\mbox{mean}_{c}) / \mbox{std}_{c})}
\end{equation}
where $\mbox{mean}_{c}$ and $\mbox{std}_{c}$ are the average and the standard deviation of the attention map of $c$-th channel, respectively.

\begin{figure}[t]
\centering
\centerline{\includegraphics[width=\columnwidth]{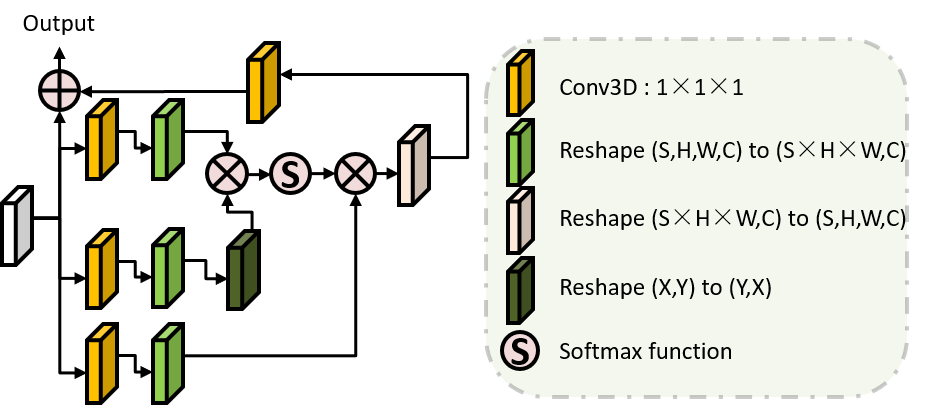}}
\caption{The structure of the non-local module. It collects global information by weighted summation of the features from all positions to the target position, where the connecting weight is calculated by the pairwise relation.  }
\label{fig:nonlocal}
\vspace{0.5cm}
\end{figure}

\subsection{Non-local Module}

The non-local module collects global information by weighting features from any location to  {every pixel point}, where the weights are calculated by pairwise relationships. 

Collecting global information captures long-range dependencies. This is often achieved by stacking convolution layers: large receptive fields can be obtained after multiple stacks of the convolution layer. Such stacking leads to inefficient computations and gradient degradation. In addition, adding a convolution layer increases parameters and makes it harder to optimize. A careful design of the optimization process is required.

Unlike the progressive behavior of recursive and convolution operations, the non-local module eliminates locality by calculating the relationship between any two locations to capture long-range dependencies rather than restricting itself to adjacent points. For convolution operations, when using a 3$\times$3 convolution operation, the output considers only the 3$\times$3 neighborhood for each location. { It is difficult to obtain the relationship between each pixel and the whole only by convolution operation.
} For nonlocal operations, the output for each location takes into account all locations rather than the immediately adjacent neighborhood. By calculating the correlation between any two locations, we can capture the long-range correlation directly. This non-local method is equivalent to constructing a convolution kernel of the same size as the feature map, so that more information can be preserved.

A convolution unit on CNN uses only the neighborhood area of the kernel size each time. Even when the size of the receptive field becomes larger at a late stage, it is still of the local nature. Because of this, the contribution of distant areas (such as long-distance pixels) to the current area is ignored. The non-local module captures this long-range relationship: for a 3D CT, there is a weight of the relationship between any pixel in all the slices and a pixel in the current slice. It is expressed as:
\begin{equation}
\mathbf{y}_{i}=\sum_{\forall j} f\left(\mathbf{x}_{i}, \mathbf{x}_{j}\right) g\left(\mathbf{x}_{j}\right)
\end{equation}
where $
f\left(\mathrm{x}_{i}, \mathrm{x}_{j}\right)=\theta\left(\mathrm{x}_{i}\right)^{T} \cdot \phi\left(\mathrm{x}_{j}\right)
$, and $\theta$, $\phi$, and ${g}$ are convolution operations with kernel size 1.

The detailed calculation process is shown in Fig.~\ref{fig:nonlocal}. A non-local module is inserted after the last attention module to make sure that the convolution follows the non-local operation. The attention mechanism in the attention module adopts mixed attention, which combines spatial attention and channel attention.

\subsection{Network Structures}
\begin{table}

\setlength{\belowcaptionskip}{0.5cm}

\label{Tab1e:Modules}
\caption{Modules used in our neural network model}
\centering 
\footnotesize{
\begin{tabular}{c|c|c|c}
\hline
Layer             & Output Size  & ResMix3        & ResMix6       \\ \hline
Conv1             & 32$\times$80$\times$80        & \multicolumn{2}{c}{7 $\times$ 7 $\times$ 7, 64, stride 2} \\ \hline
Max pooling       & 16$\times$40$\times$40        & \multicolumn{2}{c}{3 $\times$ 3 $\times$ 3 stride 2}      \\ \hline
Residual Unit     & 16$\times$40$\times$40        & \multicolumn{2}{c}{
$1\times
\begin{bmatrix}
1 \times 1 \times 1 , 64\\ 
3 \times 3 \times 3 , 64\\ 
1 \times 1 \times 1 , 256
\end{bmatrix}$
}                    \\ \hline
Attention Module  & 16$\times$40$\times$40        & attention $\times$1        & attention $\times$1       \\ \hline
Residual Unit     & 8$\times$20$\times$20        & \multicolumn{2}{c}{
$1\times
\begin{bmatrix}
1 \times 1 \times 1 , 64\\ 
3 \times 3 \times 3 , 64\\ 
1 \times 1 \times 1 , 256
\end{bmatrix}$
}                    \\ \hline
Attention Module  & 8$\times$20$\times$20        & attention $\times$1        & attention $\times$2       \\ \hline
Residual Unit     & 4$\times$10$\times$10        & \multicolumn{2}{c}{
$1\times
\begin{bmatrix}
1 \times 1 \times 1 , 64\\ 
3 \times 3 \times 3 , 64\\ 
1 \times 1 \times 1 , 256
\end{bmatrix}$
}                    \\ \hline
Attention Module  & 4$\times$10$\times$10        & attention $\times$1        & attention $\times$3       \\ \hline
Residual Unit     & 2$\times$5$\times$5          & \multicolumn{2}{c}{
$3\times
\begin{bmatrix}
1 \times 1 \times 1 , 64\\ 
3 \times 3 \times 3 , 64\\ 
1 \times 1 \times 1 , 256
\end{bmatrix}$ 
}                    \\ \hline
Average pooling   &  1$\times$1$\times$1         & \multicolumn{2}{c}{ stride 1}       \\ \hline
FC, Softmax        & \multicolumn{3}{c}{3}                                  \\ \hline
\multicolumn{2}{c|}{params$\times 10^6$} &     58.3            &          95.7      \\ \hline
\multicolumn{2}{c|}{FLOPs$\times 10^9$}  &     49.4            &          68.7      \\ \hline

\end{tabular}
}
\vspace{0.5cm}
\end{table}

The overall structure of the NL-RAN architecture is given in Table~\ref{Tab1e:Modules}. It starts with a 3$\times$3 3D convolutional layer with stride 2 and a 3D max-pooling, followed by three 3D attention modules. After that, a non-local module is used to further expand the size of receptive fields, which is then followed by a GAP and a full connection layer.

We use the cross-entropy loss function to learn the category of pneumonia prediction {In the training phase}. It mainly describes the distance between the actual output and the expected output, which is defined as follows:
\begin{equation}
H(p, q)=-\sum_{i=0}^{N} p\left(x_{i}\right) \log \left(q\left(x_{i}\right)\right),
\end{equation}
where ${N}$ is the number of categories, 
${p}$ is the probability distribution of the expected output (i.e., the real tag of the data), and ${q}$ is the probability distribution of the prediction output (i.e., the tag predicted by the model). The smaller the cross-entropy loss is, the closer the two distributions are, and the better the classification result is.

\subsection{Obtaining Attention Maps}

\begin{figure}[t]
\centering
\centerline{\includegraphics[width=\columnwidth]{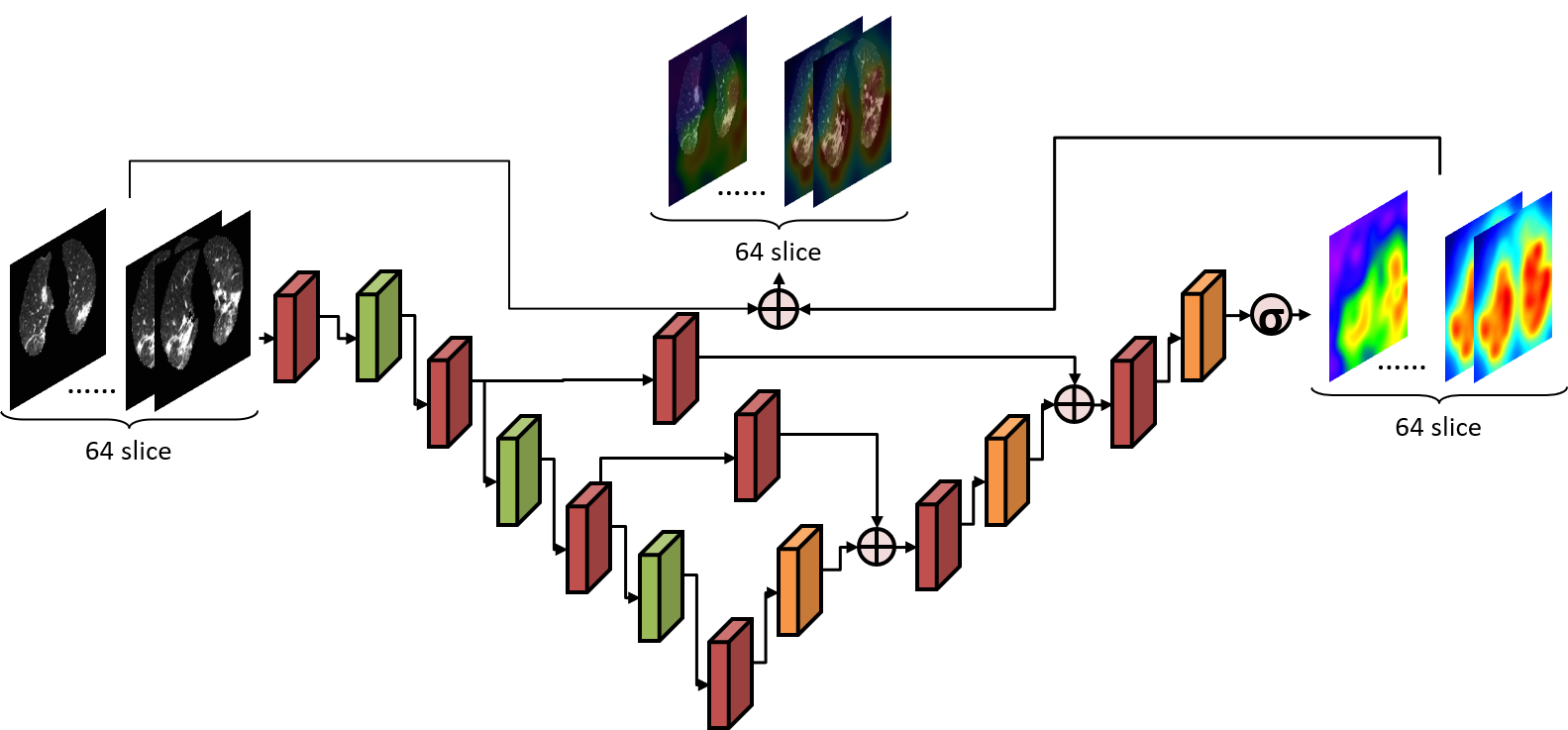}}
\caption{The architecture of our proposed 3D attention module. It takes the feature map that has been activated by the first attention module as the attention map and outputs a heat map. }
\label{fig:getAttMap}
\vspace{0.5cm}
\end{figure}

As shown in Fig.~\ref{fig:getAttMap}, our attention module is structurally similar to UNet and acts as location-based learning in our model. After several downsamplings and upsamplings of the attention module, the feature map can obtain a weight matrix with a higher weight for the lesion area. After the sigmoid activation, an attention map can be obtained. We choose the first attention module in the model and color its attention map to obtain the heat map.

We also employ the class activation mapping to increase the explainability of the proposed model on its predictions. The last convolutional layer of the classification network is followed by a global average pooling layer and a fully connected layer. Through the GAP layer, our classification model down-samples the feature size from S$\times$H$\times$W to 1$\times$1$\times$1, and thus loses the spatial representation ability.  However, the feature map before the GAP layer contains both high-level semantic and locational information. Each channel corresponds to the activation of different semantic cues. Through CAM, our model finds the response region of the prediction result.

\section{EXPERIMENTS}
\label{sec:guidelines}

\subsection{Experimental Setting}

\subsubsection{Data and Implementation Details}

{Dataset of the CT images and metadata are constructed from cohorts from the China Consortium of Chest CT Image Investigation (CC-CCII). All CT images are classified into novel coronavirus pneumonia (NCP) due to SARS-CoV-2 virus infection, common pneumonia and normal controls. This dataset is available globally with the aim to assist the clinicians and researchers to combat the COVID-19 pandemic. This dataset is provided in PNG format and is converted to a 3D DICOM format during data processing.}

We use a large public COVID-19 dataset, CC-CCII \cite{zhang2020clinically}\footnote{http://ncov-ai.big.ac.cn/download}, to evaluate the proposed method. The original CC-CCII dataset contains a total of 6,752 CT scans with 617,775 slices. However, only 4179 CT scans have been made public. Because some data failed in preprocessing, we use 4079 CT scans in our experiments. The data set contains three categories: common pneumonia (CP), novel coronavirus pneumonia (NCP), and normal.  {There are 1482 scans in CP, 1519 in NCP and 1078 in normal. When the whole data set is divided into a training set, verification set and test set according to the ratio of 8:1:1, the test set CP contains 140 scans, NCP contains 148 scans and normal contains 120 scans. For the training set, there are 1184 CP scans, 1212 NCP scans and 867 normal scans. For the verification set, there are 158 CP scans, 159 NCP scans and 91 normal scans.} To avoid the interference of non-lung tissue, we use UNet to segment the data and use the mask to extract the lung region. To preserve the {information between slices} in the model analysis phase, all CT slices from the same patient are converted into a 3D image. As the number of slices per CT scan is inconsistent, which will lead to the size of each image is not uniform. So we process each CT image.  {In this step of data processing, we process each CT slice into 160 * 160 sizes, but we do not use the down sampling method. Considering that the down sampling will cause the loss of image content, we use the clipping method to get the key content of the image, and each image is cropped with the size of 160 $\times$ 160 at a fixed position in the middle. Only in the slice dimension, we use the average interval sampling method to limit the number of slices to 64.} Less than 64 slices have 64 slices duplicated. More than 64 slices have 64 slices sampled at intervals, resulting in 64 slices per CT image. Then we randomly divide the data into the training set, validation set, and testing set according to the ratio of 8:1:1, resulting in 3,263 training images, 408 testing images, and 408 validation images. The image size is 64$\times$160$\times$160, and the pixel value range is 0 to 255.

As an image stack corresponds to a 3D volume, we use a 3D CNN for slice-by-slice segmentation. We have implemented our NL-RAN in Pytorch, and ReLU is used in our network as the nonlinear activation function. We implemented our experiments on a Ubuntu desktop with an NVIDIA RTX 2080 Ti GPU. In our experiments, the basic channel number ${C}$ in our NL-RAN was set to 64, and the downsampling and upsampling times in the attention module were 2 and 2, respectively. During training, we used the SGD optimizer and a mini-batch of 6 slices. The learning rate was initialized to 0.001, and the training was terminated when the performance on the validation set stopped to improve for 50 epochs.


\subsubsection{Evaluation Metrics}

{To evaluate our model, we use 
weighted average precision, weighted average recall, accuracy, weighted average F1-score, precision-recall (PR) curve, confusion matrix, the ROC (Receiver Operating Characteristic) curve, and the AUC (Area Under the ROC Curve) value. Weighted average metrics are widely used as evaluation metrics for classification problems \cite{korucuk2021estimation}.}


Specifically, precision and recall for class $i$ are defined as follows:
\begin{equation}
P_{i}=\frac{T P_{i}}{T P_{i}+F P_{i}}, \quad R_{i}=\frac{T P_{i}}{T P_{i}+F N_{i}}
\end{equation}
where ${TP_i}$ is the true positives of class $i$ that positive samples are judged to be positive, ${FN_i}$ is the false negatives of class $i$ that positive samples are judged to be negative, ${FP_i}$ is the false positives of class $i$ that negative samples are judged as positive, and ${TN_i}$ is the true negatives of class $i$ that negative samples are judged as negative. $P_{i}$ and $R_{i}$ represent precision and recall of class $i$, respectively. Recall reflects the ability of a classifier to cover positive examples, and precision reflects the accuracy of a classifier to predict positive examples.

{
Suppose $W_i:=\frac{T P_{i}+F N_{i}}{\sum_{j=1}^{N}\left(T P_{j}+F N_{j}\right)}$ is the weight of the $i$-th category. Then the weighted average precision $P$ can be expressed as 
}
{
\begin{equation}
\begin{aligned}
P &= \sum_{i=1}^{N} W_{i} * P_{i} \\
  &= \sum_{i=1}^{N}\left(\frac{T P_{i}+F N_{i}}{\sum_{j=1}^{N}\left(T P_{j}+F N_{j}\right)} * \frac{T P_{i}}{T P_{i}+F P_{i}}\right).  
\end{aligned}
\end{equation}
}
{
Similarly, the weighted average recall $R$ can be obtained by
}
{
\begin{equation}
\begin{aligned}
R&= \sum_{i=1}^{N} W_{i} * R_{i}\\
&=\sum_{i=1}^{N}\left(\frac{T P_{i}+F N_{i}}{\sum_{j=1}^{N}\left(T P_{j}+F N_{j}\right)} * \frac{T P_{i}}{T P_{i}+F N_{i}}\right)\\
&=\sum_{i=1}^{N}\left(\frac{T P_{i}}{\sum_{j=1}^{N}\left(T P_{j}+F N_{j}\right)}\right)\\
&=\frac{\sum_{i=1}^{N} T P_{i}}{\sum_{j=1}^{N}\left(T P_{j}+F N_{j}\right)}
\end{aligned}
\end{equation}
}
{
In multi-class, accuracy (ACC) is the number of classified correctly divided by the total number. The corresponding numerator should be the sum of TP for each category, and the denominator should be the sum of TP and FN for each category. Therefore,
}
{
\begin{equation}
ACC=\frac{\sum_{i=1}^{N} T P_{i}}{\sum_{j=1}^{N}\left(T P_{j}+F N_{j}\right)} \\
\end{equation}
}
{
Comparing recall R and accuracy ACC, we can find the value of R equals to the ACC when using the weighted average method. However, in order to ensure the completeness and diversity of the evaluation, we report both of their values.
}

{The weighted average F1 is calculated by the weighted average Precision $P$ and the weighted average Recall $R$. The formula is as follows.}
{
\begin{equation}
\begin{aligned}
F_{1}&=\frac{2}{1 / P+1 / R}\\
&=\frac{2}{\frac{\sum_{j=1}^{N}\left(T P_{j}+F N_{j}\right)}{\sum_{i=1}^{N} \frac{T P_{i}\left(T P_{i}+F N_{i}\right)}{T P_{i}+F P_{i}}}+\frac{\sum_{j=1}^{N}\left(T P_{j}+F N_{j}\right)}{\sum_{i=1}^{N} T P_{i}}}
\end{aligned}
\end{equation}
}



\begin{figure*}[t]
\centering
\centerline{\includegraphics[width=\textwidth]{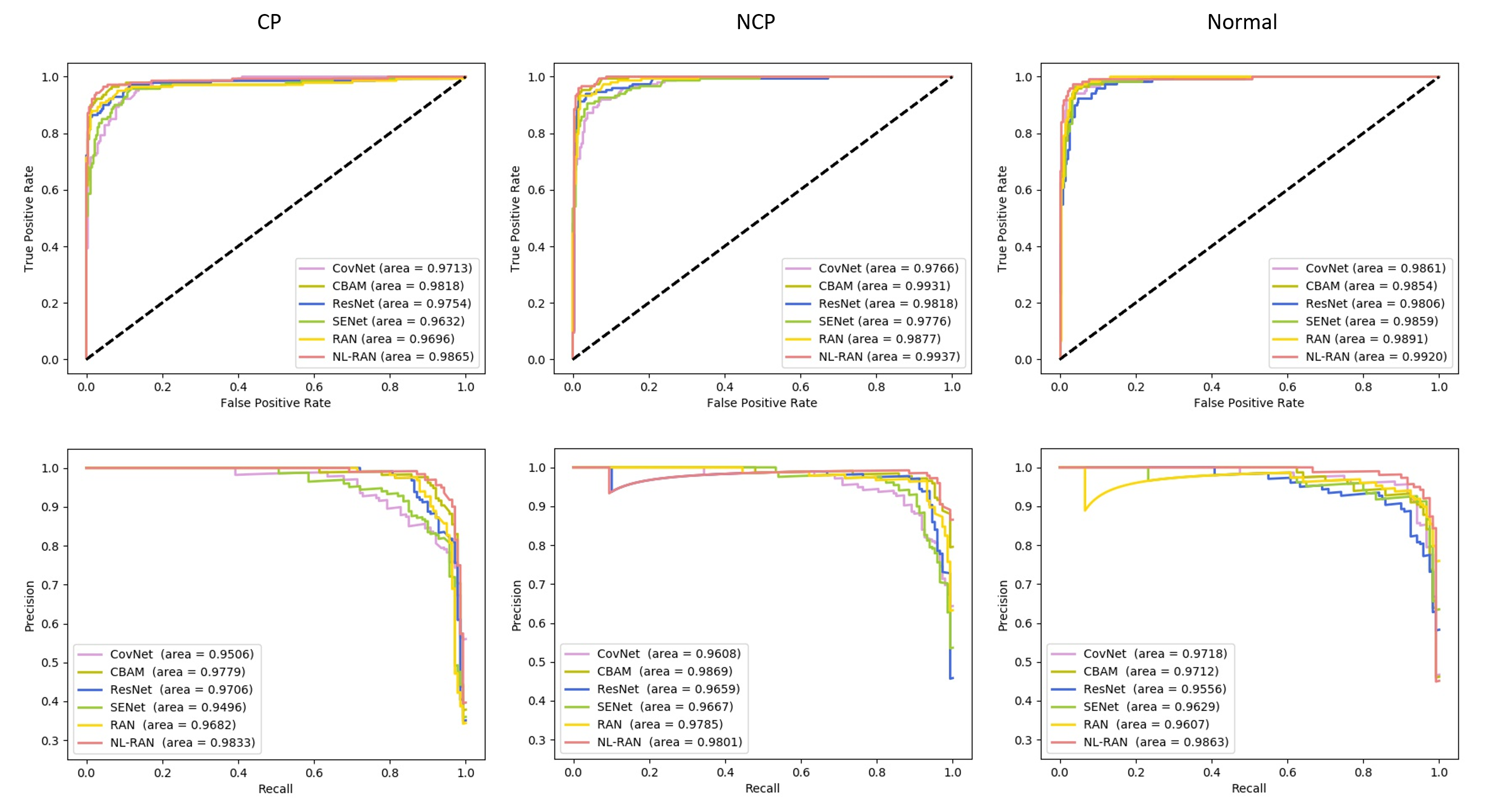}}
\caption{Comparison of six methods on three-class ROC and PR curves. The larger the area enclosed by the curve, the better the performance of the model. The classification of common pneumonia (CP) is in the left column, COVID-19 (NCP) in the middle column, and normal in the right column.}
\label{ROC_PR}
\vspace{0.5cm}
\end{figure*}
The ROC curve is obtained by plotting the True Positive Rate (TPR) against the False Positive Rate (FPR). TPR, also known as recall, represents the percentage of actual positive samples that are correctly predicted by the classifier. FPR represents the percentage of actual negative samples that are falsely predicted as positive by the classifier. Therefore, the closer the ROC curve is to the point (0, 1), the better the classifier prediction will be. TPR and FPR are defined as follows:
\begin{equation}
T P R_{i}=\frac{T P_{i}}{T P_{i}+F N_{i}}, \quad F P R_{i}=\frac{F P_{i}}{F P_{i}+T N_{i}}
\end{equation}
where ${T P R_{i}}$ and ${F P R_{i}}$ represent the TPR value and FPR value corresponding to the ROC curve of class $i$, respectively.


\subsection{Comparison with Existing Works}
We compare our proposed NL-RAN with the following most advanced networks for image classification: 1) CBAM, which extends ResNet by combining channel attention and spatial attention and achieves the top performance in several natural image classification tasks; 2) SENet, which uses channel attention to enable the network to focus more on channels; 3) RAN, which combines stackable attention module; 4) ResNet, an efficient and universal CNN that achieves the most advanced image classification performance; 5) CovNet, a concise and efficient CNN for identifying pneumonia lesions that has achieved a good performance in the classification of lesions. We use 3D versions of CBAM, SENet, RAN, and ResNet, and set their channel numbers in the same way as in our NL-RAN, i.e., 64 in the first module and doubled after each down-sampling. Two versions of our NL-RAN are used in our experiments, NL-RAN3 and NL-RAN6, which stack the attention module three times and six times, respectively. These networks were trained separately with the same data sets.

\begin{table*}[t]
\centering
\caption{\textit{ \small  Performance of different networks on the testing set. The same backbone is used for these networks. The weighted average method is used to calculate the average indicator of three categories. 
} }
\resizebox{1.8\columnwidth}{!}{
\begin{tabular}{c|ccccc}
\hline
\textbf{Method}        &\textbf{ACC} ($\%$)  & \textbf{P} ($\%$) & \textbf{R} ($\%$) & \textbf{F1} ($\%$) &  \textbf{AUC}($\%$)   \\ \hline 
CovNet  \cite{li2020artificial}& 92.108±0.636  & 92.306±0.563  & 92.108±0.636  & 92.104±0.659  & 98.166±0.376     \\
CBAM  \cite{woo2018cbam}   & 92.502±0.975  & 92.736±0.867  & 92.502±0.975    & 92.526±0.967     & 98.110±0.633      \\
ResNet  \cite{he2016deep}  & 92.698±0.699   & 92.762±0.748 & 92.698±0.699   & 92.702±0.715     & 96.476±0.819       \\
SENet  \cite{hu2018squeeze}& 92.744±0.876   & 92.846±0.948 & 92.744±0.876  & 92.750±0.885    & 98.134±0.440       \\
RAN  \cite{wang2017residual}& 93.432±0.470   & 93.568±0.419  & 93.432±0.470 & 93.450±0.467   & 98.134±0.504       \\\hline
NL-RAN & \textbf{94.264±0.446}  & \textbf{94.370±0.405}    & \textbf{94.264±0.446} & \textbf{94.274±0.439}   & \textbf{98.970±0.108}  \\ \hline
\end{tabular}
}
\label{table:Acc}
\end{table*}

Fig. \ref{ROC_PR} shows both the ROC curve and the PR curve of these networks for each category. We can see that CBAM achieves the lowest AUC in the pneumonia lesion classification task. This can be explained that CBAM is designed for learning from large-scale RGB images, which have different features and intensity distributions from our CT images. SENet has a higher AUC than ResNet, mainly because SENet has embedded the channel attention module, which makes the network more focused on channels. Compared with 3D RAN, our NL-RAN achieves the highest AUC. We conclude that our NL-RAN achieves the best performance among all these networks.

Table~\ref{table:Acc} shows the performance in terms of ACC, precision (P), recall (R), F1-score, and AUC of comparing networks. CovNet performs worse than 3D ResNet because the latter analyzes CT directly as 3D input while the former analyzes CT slices one by one, with the information between slices ignored. As a result, we adopt a 3D CNN to predict the classification label for each CT image in our proposed network. 


In addition, the metrics of all methods are high and the difference between them seems to be very small. It is not clear if the difference observed is due to a systematic increase in performance by the proposed method or due to randomization of network weights. So it is necessary to do sensitivity analysis. We report the average metrics performance by using five random weight initialization to run our method and the compared methods in Table~\ref{table:Acc}. 
{From Table \ref{table:Acc}, we can see that our method outperforms compared methods. 
Similarly, in order to clarify that the difference between these metrics does not come from the influence of finite sample size, we did some experiments on the training sample size. The size of training samples are 80\%, 70\%, 60\%, 50\%, 40\% of the total number of data samples. We report the results in Table~\ref{table:sizeData}. From the table, we can find the fewer samples participate in the training, the worse the test results. However, our model still achieves the best results in all different sample size settings. In summary, we conclude that the difference between these models is not due to variability with respect to neural network weight initialization or finite sample size variability.}


\begin{figure*}[t]
\centering
\centerline{\includegraphics[width=\textwidth]{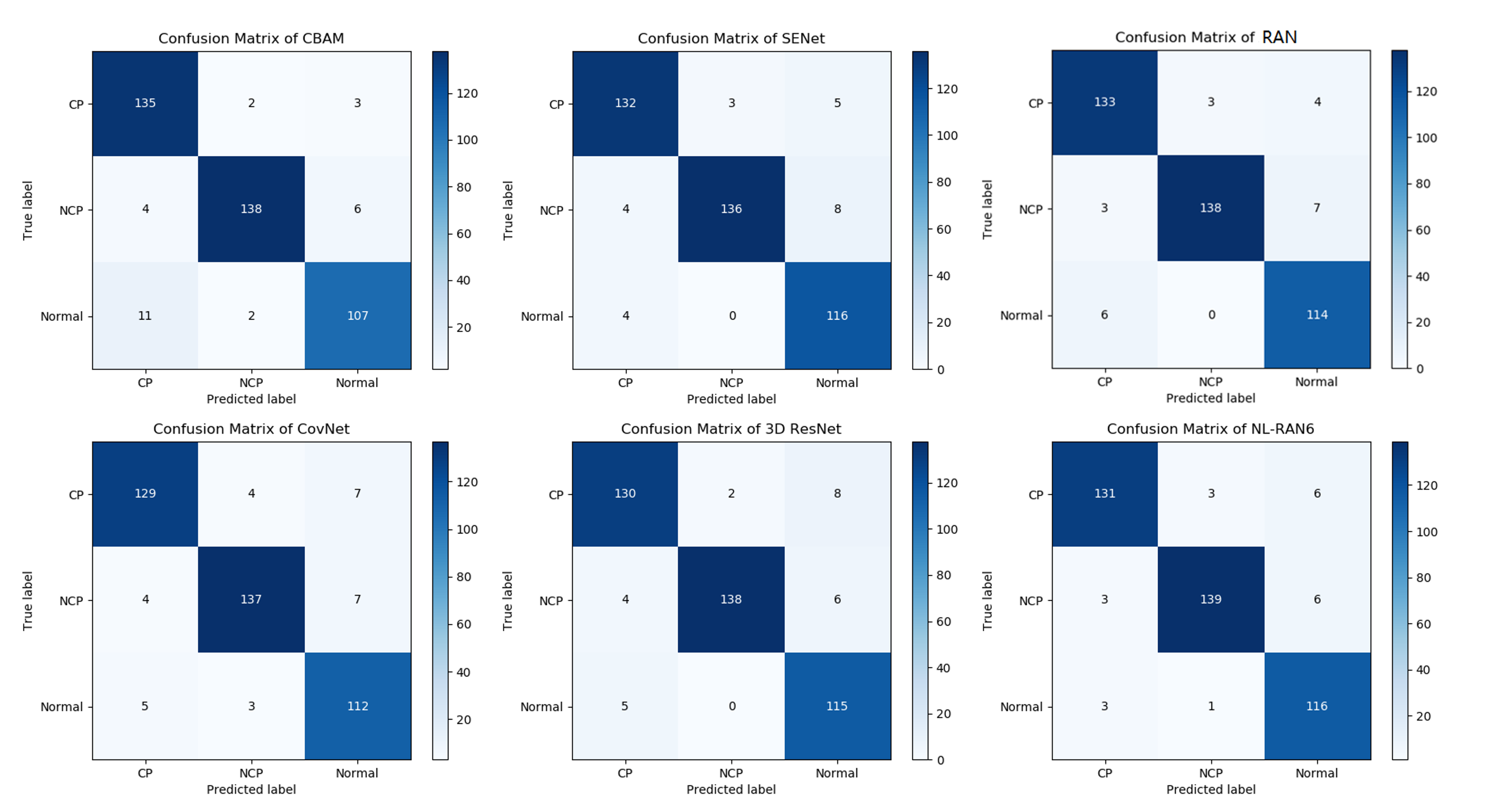}}
\caption{Confuse matrix of our proposed method and mainstream methods.}
\label{mtxs}
\end{figure*}

Compared with 3D ResNet, 3D SENet and CBAM embed two different attention mechanisms to focus on critical areas. Table~\ref{table:Acc} shows that SENet performs better than CBAM but 3D RAN performs better than both. This agrees with our ablation study results, which indicate that embedding spatial attention in a 3D model is not effective but that mixed attention that combines spatial attention and channel attention enables a model to effectively learn key features.

Our proposed method uses 3D CNN combined with the mixed attention module to focus the model on critical areas. It uses residual links to effectively learn while embedding the non-local module to increase the size of receptive fields to focus on global information. With several attention modules and non-local, our network requires more computations than methods using a single CNN. Nevertheless, on GPU2080Ti, our method can predict 400 samples in less than 15 seconds, making it practical to  assist diagnosis in real time.

The confusion matrix in our proposed method is shown in Fig.~\ref{mtxs}, along with those of CovNet and 3D ResNet, where the $x$-axis and $y$-axis are the predicted labels and the ground truth labels of testing CT images, respectively. In the CT images, the appearance of COVID-19 lesions is not unique. Some lesions are similar to common pneumonia lesions. As a result, COVID-19 and common pneumonia are easily  misclassified into each other.

As shown in Fig.~\ref{mtxs}, CovNet has the most misclassifications while our method has the least. In CovNet's confusion matrix, there are seven samples with the true label as NCP and predicted as Normal, whereas three samples with the true label as Normal and predicted as NCP, that is, there are 10 samples misclassified between NCP and Normal categories. In 3D ResNet, only six samples were misclassified between NCP and Normal because CovNet analyzed each slice of CT scan, and 3D ResNet analyzed a CT scan as a whole, taking into account time-series information, so the importance of establishing a 3D model for lesion classification can be seen. However, in 3D ResNet, there are five samples with the true label as CP and predicted as Normal, whereas eight samples with the true label as Normal and predicted as CP, that is, 13 samples were misclassified between NCP and Normal categories. In NL-RAN6, only nine samples were misclassified between the CP and Normal categories, because NL-RAN6 embedded attention module  and non-local module, which could focus on the details of lesions while focusing on global information, reduced the number of misclassified samples.

\begin{table}[t]
\setlength{\belowcaptionskip}{0.5cm}

\centering
\caption{\textit{ \small  {In several groups of experiments, we set the training sample size as 80$\%$, 70$\%$, 60$\%$, 50$\%$ and 40$\%$ of the total sample size. In each experiment, our model can still maintain its superiority. }} }
\resizebox{\columnwidth}{!}{

\begin{tabular}{c|c|ccccc}
\hline
\textbf{Size}    &\textbf{Method}    &\textbf{ACC} ($\%$)  & \textbf{P} ($\%$) & \textbf{R} ($\%$) & \textbf{F1} ($\%$) &  \textbf{AUC}($\%$)   \\ \hline 
\multirow{6}{*}{80\%}           & CovNet          & 92.4           & 92.69          & 92.4           & 92.42          & 97.33          \\
              & CBAM            & 92.65          & 92.83          & 92.65          & 92.69          & 98.42          \\
             & ResNet          & 92.65          & 92.94          & 92.65          & 92.69          & 98.72          \\
              & SENet           & 93.38          & 93.57          & 93.38          & 93.41          & 97.96          \\
              & RAN             & 93.87          & 93.95          & 93.87          & 93.89          & 98.2           \\ \cline{2-7} 
              & NL-RAN          & \textbf{94.85} & \textbf{94.88} & \textbf{94.85} & \textbf{94.85} & \textbf{98.9}  \\ \cline{1-7} 
\multirow{6}{*}{70\% }          & CovNet          & 91.42          & 91.69          & 91.42          & 91.39          & 97.76          \\
              & CBAM            & 92.4           & 92.54          & 92.4           & 92.41          & 98.07          \\
              & ResNet          & 92.16          & 92.18          & 92.16          & 92.14          & 95.67          \\
              & SENet           & 92.65          & 92.65          & 92.65          & 92.61          & 97.77          \\
              & RAN             & 93.63          & 93.74          & 93.63          & 93.65          & 98.77          \\ \cline{2-7} 
              & NL-RAN          & \textbf{94.61} & \textbf{94.73} & \textbf{94.61} & \textbf{94.62} & \textbf{99.03} \\ \cline{1-7} 
\multirow{6}{*}{60\%}           & CovNet          & 90.93          & 91.21          & 90.93          & 90.97          & 97.81          \\
              & CBAM            & 91.91          & 92.29          & 91.91          & 91.94          & 97.13          \\
              & ResNet          & 91.18          & 91.53          & 91.18          & 91.21          & 97.85          \\
              & SENet           & 92.16          & 92.31          & 92.16          & 92.19          & 97.34          \\
              & RAN             & 93.38          & 93.52          & 93.38          & 93.38          & 98.36          \\ \cline{2-7} 
              & NL-RAN          & \textbf{93.87} & \textbf{94.04} & \textbf{93.87} & \textbf{93.89} & \textbf{98.92} \\ \cline{1-7} 
\multirow{6}{*}{50\%}           & CovNet          & 90.2           & 90.27          & 90.2           & 90.13          & 98.06          \\
              & CBAM            & 91.18          & 91.53          & 91.18          & 91.21          & 97.85          \\
              & ResNet          & 90.44          & 90.6           & 90.44          & 90.47          & 97.64          \\
              & SENet           & 91.91          & 91.97          & 91.91          & 91.92          & 98.02          \\
              & RAN             & 92.16          & 92.29          & 92.16          & 92.17          & 98.33          \\ \cline{2-7} 
              & NL-RAN          & \textbf{92.89} & \textbf{93.03} & \textbf{92.89} & \textbf{92.91} & \textbf{98.64} \\ \cline{1-7} 
\multirow{6}{*}{40\%}           & CovNet          & 86.03          & 86.51          & 86.03          & 86.09          & 95.56          \\
              & CBAM            & 89.22          & 89.4           & 89.22          & 89.27          & 96.7           \\
              & ResNet          & 88.48          & 89.09          & 88.48          & 88.19          & 95.34          \\
              & SENet           & 90.93          & 91.21          & 90.93          & 90.97          & 97.81          \\
              & RAN             & 90.2           & 90.34          & 90.2           & 90.22          & 98.31          \\ \cline{2-7} 
              & NL-RAN          & \textbf{91.91} & \textbf{91.96} & \textbf{91.91} & \textbf{91.9}  & \textbf{98.74} \\ \cline{1-7} 
\end{tabular}
}
\label{table:sizeData}
\end{table}

\subsection{Ablation study}

\subsubsection{Effectiveness of the Attention Module}
We performed ablation experiments and sensitivity analysis on all modules, as shown in Table \ref{table:Ablation}. We first evaluate different combinations of 3D channel attention and 3D spatial attention in the ResNet backbone. We use the same training set to train three different attention networks. ResCA and ResCSA in Table \ref{table:Ablation} show the classification performance on the test set with direct stacking of attention modules. Note that ResCSA stacks 3D spatial attention as well as 3D channel attention. Its performance is not better than that of ResCA, which stacks only 3D channel attention. It can be observed that the classification performance deteriorates as the stacked attention module structure becomes too complex compared with ResNet. Therefore, in the following experiments, we reduced the number of stacks of attention modules. 



We further combine 3D channel attention and 3D spatial attention directly with the ResNet backbone, that is each attention module is stacked only once in the backbone. Table \ref{table:Ablation} shows the classification performance of the three attention modules directly in combination with ResNet. That is to embed channel attention(CA), spatial attention(SA), or channel spatial attention(CSA) in ResNet. It can be seen that the direct combination of 3D attention embedded in ResNet provides better performance than the ResCSA. However, we can find 3D spatial attention does not play the role as we have expected. As a result, we explore another mixed attention method in the following experiments.

\begin{table*}[t]
\setlength{\belowcaptionskip}{0.5cm}

\centering
\caption{\textit{\small Result of Ablation Experiment. ResNet is used as the backbone network.}}
\resizebox{1.8\columnwidth}{!}{
\begin{tabular}{c|ccccc}
\hline
\textbf{Method} &\textbf{ACC} ($\%$)  & \textbf{P} ($\%$) & \textbf{R} ($\%$) & \textbf{F1} ($\%$) &  \textbf{AUC} ($\%$)     \\ \hline
ResCSA          & 90.146±2.314          & 90.462±2.222          & 90.146±2.314          & 90.162±2.286          & 97.340±0.995          \\
SA              & 90.786±1.494          & 91.032±1.536          & 90.786±1.494          & 90.820±1.478          & 97.290±0.641          \\
CSA             & 91.374±1.004          & 91.570±1.016          & 91.374±1.004          & 91.396±0.997          & 97.504±1.118          \\
CA              & 91.503±0.635          & 91.703±0.554          & 91.503±0.635          & 91.537±0.629          & 97.437±0.818          \\
ASPP            & 91.547±0.953          & 91.693±0.951          & 91.547±0.953          & 91.562±0.946          & 98.317±0.349          \\
ResCA           & 91.570±0.823          & 91.710±0.805          & 91.570±0.823          & 91.570±0.819          & 98.242±0.349          \\
ResMix6         & 91.962±1.140          & 92.092±1.157          & 91.962±1.140          & 91.974±1.142          & 98.350±0.302          \\
ResMix3         & 92.060±1.048          & 92.216±1.057          & 92.060±1.048          & 92.078±1.047          & 98.202±0.393          \\\hline
NL-ResMix3      & 92.940±0.922          & 93.070±0.967          & 92.940±0.922          & 92.952±0.933          & 98.684±0.296          \\ 
NL-ResMix6      & \textbf{93.872±0.627} & \textbf{94.004±0.614} & \textbf{93.872±0.627} & \textbf{93.886±0.622} & \textbf{98.928±0.183} \\ \hline
\end{tabular}
}
\label{table:Ablation}
\vspace{0.5cm}
\end{table*}

RAN introduces a 2D attention module. The attention module adopts a bottom-up top-down structure. First, a series of convolution and pooling are adopted to gradually extract the features of the upper layer and increase the size of the receptive fields of the model. The same number of upsampling is used to enlarge the feature map to the same size as the original input to get the attention map of the input. 
The bottom-up top-down encoder-decoder structure is commonly used in image segmentation.
It is equivalent to a weakly-supervised positioning of the task of learning.
With this structure, the attention module can effectively focus on the detail of an image. 
We borrow the idea and construct a similar attention module.

However, stacking the attention module directly results in significant performance degradation because the range of the attention map after activation is [0, 1] and repeated multiplication gradually reduces the eigenvalues. As a result, the performance of ResCSA is inferior to other attention networks. To solve this problem, we adopt the stacking of the attention module in RAN: in our 3D mixed attention, the attention map and feature map are multiplied together with the original feature map. It can ensure that the attention map does not affect subsequent calculations when its values approach 0. Mixed attention is achieved with a simple sigmoid for each channel and space location.

Table \ref{table:Ablation} shows the model performance with the 3D mixed attention module embedded. ResMix3 means that the mixed attention module is superimposed three times. From the table we can be see that there is little difference in performance between the models that stack the mixed attention module three times and six times(ResMix6), respectively, which indicates that stacking attention modules can no longer effectively expand the size of receptive fields. Comparing with ResCSA, ResCA, and ResMix3, we can see that the 3D mixed attention performs better than these models, which indicates the advantages of the 3D mixed attention in paying attention to image details.

\subsubsection{Effectiveness of the Non-local Module}
We go one step further by adding non-local to improve our 3D RAN. As we described before, non-local operations capture long-range dependencies directly by calculating the interactions between any two locations rather than being limited to adjacent points, which is equivalent to creating a convolution kernel of the size of the feature map to maintain more information. Our above experiments (ResMix3 and ResMix6 in Table \ref{table:Ablation}) indicate that the stacked attention module cannot improve the performance, even though it expands the size of the receptive fields. Non-local, on the other hand, can augment the receptive field directly without stacking it. We insert the non-local module behind the last attention module of 3D RAN(NL-ResMix3 and NL-ResMix6). {
The results are shown in table~\ref{table:Ablation}. Comparing ResMix3 and NL-ResMix6 in the table, we can see that after embedding the non-local module into the 3D RAN, the accuracy is significantly improved. The table shows non-local combined with our 3D RAN achieves the best accuracy for the entire model on the test set.} In cases where the convolution layer cannot be stacked, using non-local can effectively increase the receptive field and improve the model performance.

\subsubsection{Effectiveness of all Components}

\begin{figure}[t]
\setlength{\belowcaptionskip}{0.5cm}

\centering
\centerline{\includegraphics[width=0.95\columnwidth]{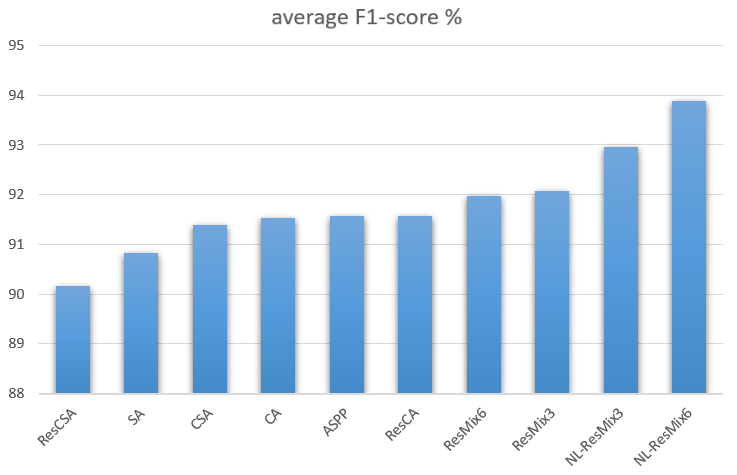}}
\caption{The F1-score of the ablation studies of the proposed method by considering different combinations of attention module. }
\label{avgF1}
\vspace{0.5cm}
\end{figure}


Our proposed NL-RAN network can combine different types of attention modules in an end-to-end trainable way, ablation of different combinations of attention modules is studied. The modules we compare include channel attention (CA), spatial attention (SA), channel spatial attention  (CSA), residual block embedded CA (ResCA) and CSA (ResCSA), ASPP for enlarging receptive fields, mixed attention mechanism  (ResMix3), and ResMix with non-local (NL-ResMix6). 
Two versions of ResMix and NL-ResMix are used, denoted by the ending number, with ResMix stacked 3 times and 6 times, respectively. Table~\ref{table:Ablation} shows their performance 
and Fig.~\ref{avgF1}  shows F1-scores of the ablation study. When receptive fields are enlarged by stacking attention modules, the F1-score increases, and ResMix3 achieves the highest F1-score without using non-local. These results imply that mixed attention can better represent features but is not suitable for enlarging receptive fields. We can also see that ResMix with embedded non-local module(NL-ResMix6) has the highest F1-score. It is worth noting that recall results become better with the increasing number of mixed attention module. 

\begin{figure*}[t]
\centering
\centerline{\includegraphics[width=\textwidth]{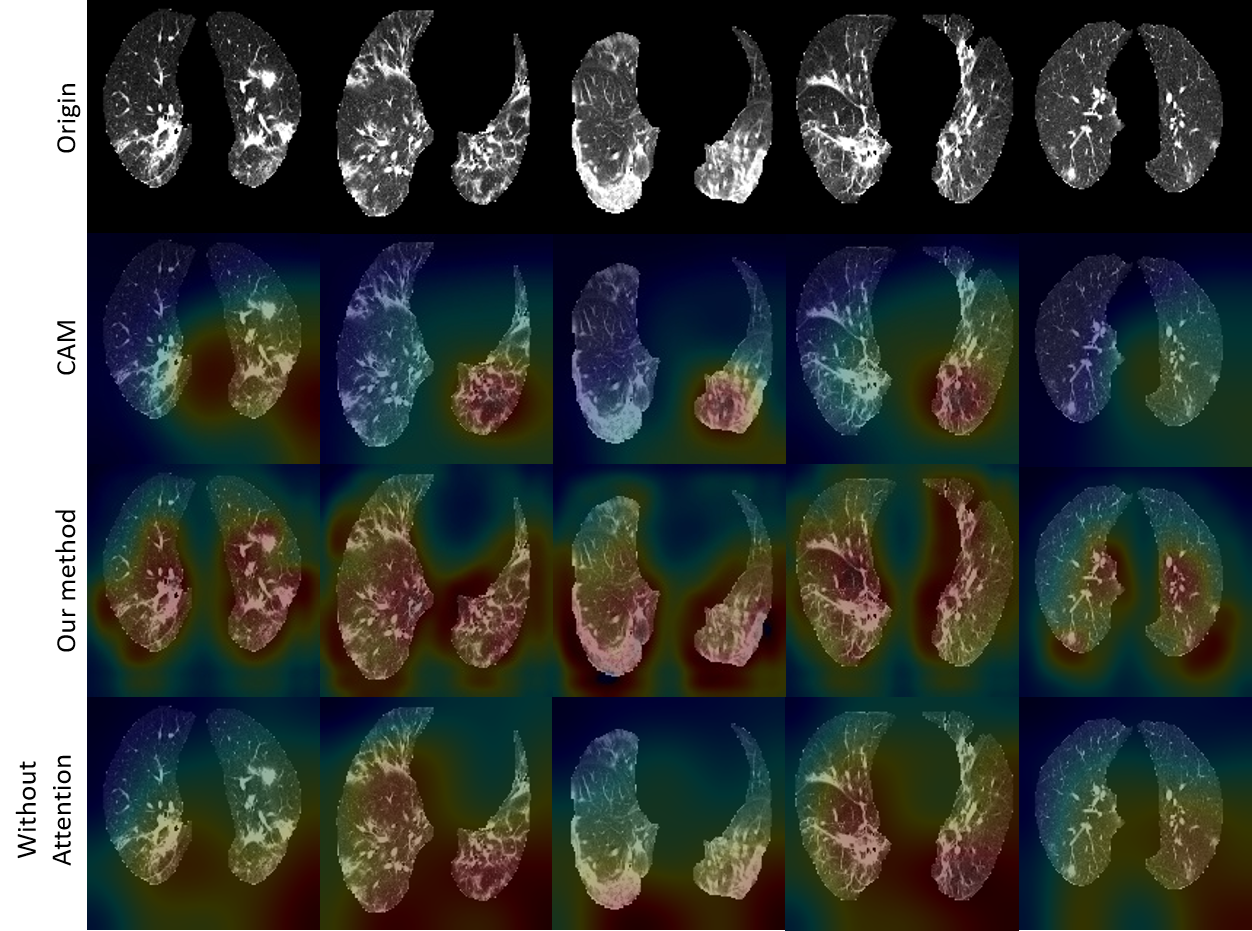}}
\caption{Heat map derived from the proposed neural network by CAM and our method. The heat map inferred by CAM is less accurate than ours due to its lack of contour information. {Because the CAM method takes a feature map at the end of the network as input, the feature map at this time has lost its outline information due to multilayer convolution.} }
\label{Visual}
\vspace{0.5cm}
\end{figure*}

\subsection{Explanation by Visualization}

We selected some data from three categories in the test set for visual analysis. To ensure that the results of visual analysis are not accidental, these data were randomly selected and were not trained.

We compare the heat map using the CAM output with the heat map of our attention module output following the process shown in Fig.~\ref{fig:getAttMap}. Fig.~\ref{Visual} illustrates a comparison of five NCP categories of lung CT for visualization heat maps. The second row shows the class activation mapping heat map of our classification model, and the last row shows the heat map of the attention module in our classification model.

It can be seen from the image that the heat map of CAM does not fully highlight the lesion area, some of which cover the lesion area but also cover too many normal tissues. NCP lesions are irregularly shaped and unevenly distributed, and CAM heat maps are derived from the last convolution layer in the model that has lost contour information. That is, the heat map obtained by CAM can only refer to the location of the lesion, but cannot provide an accurate contour. So the heat map of the CAM is not completely accurate. Looking at the heat maps output by our attention module, we can see that almost all of the lesions, whether small or obvious, are highlighted by the heat maps, even if some are widely distributed. The areas where lesions are concentrated are clearly highlighted by heat maps, the sparse areas are slightly highlighted, and the lung tissue without lesions is not highlighted. This is because the attention module serves as a feature selector to enhance good features and suppress invalid ones. These lesions are enhanced by our attention module as features that determine the image category, so our thermogram contains more accurate lesion areas. {At the same time, our ablation study compared the heat map output without Attention module. From the last line in the graph, we can see that the highlighted area of heat map has little direct relationship with the lesion area in the picture when Attention module is not used.} So we conclude that the heat map output from our attention module is more accurate than that from the CAM method. It also demonstrates that our attention module effectively enhances the characteristics of lesion areas and plays a positive role in model analysis.

\section{Conclusion}

In this paper,  we propose a novel deep 3D attention non-local network to solve the pneumonic lesions classification problem. Compared with existing deep learning methods,  the proposed method can simultaneously learn better deep feature representations from local detailed information and global information to achieve better classification accuracy.  Our experimental results indicate that our proposed method performs significantly better than existing methods. In addition, the first attention module outputs a heat map containing detailed outline information to increase the interpretability of the model.  Our heat map provides a more sensitive and accurate visualization to the infected area than CAM. Our experiments indicate that the inference of our model is fast. It can provide real time assistance of diagnosis. 

Since the model proposed in this paper is a 3D model and the required datasets must also be processed in the corresponding 3D format, the time complexity of model training is higher than that of the general model. The training of 3263 data (about 20832 slices) on GPU2080Ti in this paper takes about 24 hours. The model takes approximately 15 seconds to infer 400 data (about 25600 slices) and is able to infer faster on graphics cards with better capacity and quality. In addition, accuracy and F1 scores depend on prevalence, but the prevalence of different datasets is uncertain, so the accuracy of this article cannot be replicated to other different datasets for reference. However, the methods, ideas and models in this paper can be migrated to other fields, such as classification of medical MRI and classification of meteorology.



\ifCLASSOPTIONcaptionsoff
  \newpage
\fi



\bibliographystyle{IEEEtran}
\bibliography{IEEEabrv, bibtex/bib/example}

\end{document}